# Self-field and magnetic-flux quantum mechanics

Paul Harris
lagoonscience@worldnet.att.net

Self-field and quantized magnetic-flux are employed to generate the quantum numbers $n$, $m$ and $\ell$ of atomic physics. Wave-particle duality is shown to be a natural outcome of having a particle and its self-field.

First consider a ball bearing held between two infinite walls by two mass less springs as an example of a particle and its self-field. For the ball bearing mass $m$

$$m\ddot{x} = -k(x - x_0) \tag{1}$$

with $x_0$ being the equilibrium position, $k$ the spring constant, and the overhead dot denotes total time derivative. Equation (1) has the solution

$$x = x_0 + \xi_0 \cos \omega t, \qquad \xi \equiv x - x_0 \tag{2}$$

with $\xi_0$ denoting the displacement (relative to the equilibrium position) at t=0. Thus

$$m\dot{x} = -m\omega \xi_0 \sin \omega t, \tag{3a}$$

$$m\ddot{x} = -m\omega^2 \xi_0 \cos \omega t. \tag{3b}$$

With the interpretation

$$\dot{p}_x^{(p)} = m\ddot{x}, \qquad \dot{p}_x^{(s)} = m\omega^2 \xi_0 \cos \omega t, \tag{4}$$

Eqs (4) then satisfy

$$\dot{p}_x^{(p)} + \dot{p}_x^{(s)} = 0 \tag{5}$$

which is identical to Eq. (1). Eq. (5) can be taken as the heart of self-field (or electromagnetic reaction[1,2]) physics. It is what is expected in the absence of external forces. If Eqs. (4) are substituted into Eq. (5) and then integrated over time

$$p_x^{(p)} + p_x^{(s)} = m\dot{x}_0 = \text{constant} \tag{6}$$

so that $\dot{x}_0 = 0$ yields

$$p_x^{(p)} + p_x^{(s)} = 0. \tag{7}$$

The self-field is extended in space (distributed throughout the spring). Ignoring the possibility that the collapse process might introduce external fields, we see that the particle can not be captured until we also capture (i.e. collapse) the self-field (spring motion).

## SELF-FIELD FOR AN ELECTRON ORBIT

An electromagnetic system composed of a charge $q$ particle and interacting vector potential field has a canonical momentum $\boldsymbol{P}$ given by[3]

$$\boldsymbol{P} = \boldsymbol{p} + qc^{-1}\boldsymbol{A}. \tag{8}$$

If $\boldsymbol{P}=0$ (corresponding to a total system momentum of zero) then there is agreement with the simple Eq. (7) result providing that

$$\boldsymbol{p}^{(s)} = qc^{-1}\boldsymbol{A}^{(s)}. \tag{9}$$

For a planar closed orbit Eq. (9) results in

$$\oint p_\varphi^{(s)} r d\varphi = \frac{q}{c}\oint A_\varphi^{(s)} r d\varphi = \frac{q}{c}\Phi_\varphi^{(s)} \tag{10}$$

where $\Phi_\varphi^{(s)}$ is the magnetic self-flux linking the orbit $\boldsymbol{\varphi}$ (it does not include flux associated with an external magnetic field).

Now for a simplification and the first crucial assumption; orbits are taken as circular ($r = R$) and $A_{\varphi,\varphi}^{(s)} = 0$. That is $A_\varphi^{(s)}$ is uniform on the orbit. As a consequence Eqs. (9) and (10) yield

$$A_\varphi^{(s)} = (2\pi R)^{-1}\Phi_\varphi^{(s)}, \qquad p_\varphi^{(s)} = q(2\pi Rc)^{-1}\Phi_\varphi^{(s)}, \tag{11}$$

so that for an s-state orbit (i.e. $P_\varphi = 0$, as $RP_\varphi$ is the particle and self-field system angular momentum)

$$p_\varphi^{(p)} = -p_\varphi^{(s)} = e(2\pi Rc)^{-1}\Phi_\varphi^{(s)}, \tag{12}$$

where e is the electron charge.

The second crucial assumption is

$$\Phi_\varphi^{(s)} = n_\varphi h c e^{-1}, \quad n_\varphi = 1,2,3,..., \tag{13}$$

for the self-flux interior to a closed $\varphi$ circuit current. Eq. (13) is reminiscent of the quantized flux associated with a persistent (superconducting) current. This should not be surprising as a stable atomic orbit is certainly a "persistent" current.

From Eqs. (12) and (13)

$$p_\varphi^{(p)} = -p_\varphi^{(s)} = n_\varphi h (2\pi R)^{-1} \tag{14}$$

so that $n_\varphi$ will be provisionally recognized as the principal quantum number of conventional quantum mechanics (which is further justified by the energy expression of Eq. (17)), and the other quantum numbers will be identified in the next section. Note that if $R \alpha n_\varphi^2$, as for a Bohr atom, then $p_\varphi^{(s)}$ and $p_\varphi^{(p)}$ vary as $n_\varphi^{-1}$. Thus the momenta are expected to vanish, while, $\Phi_\varphi^{(s)} \to \infty$ as $R \to \infty$.

The relativistic expression for the energy, $E$, of an orbiting charge $q$ in the presence of the electrostatic potential $\Omega$ (infinite mass nucleus) is given by[4]

$$(E - q\Omega)^2 = (c\,\boldsymbol{P} + c\,\boldsymbol{p}^{(p)})^2 + m^2 c^4, \tag{15}$$

where $\boldsymbol{P}$ is the canonical (total) momentum of Eq. (8).

Expanding Eq. (15) to lowest order in field and momentum quantities, and again letting $q = -e$ yields, again for s-states,

$$E \approx mc^2 + (2m)^{-1} \left(p_\varphi^{(p)}\right)^2 - e^2 R^{-1}. \tag{16}$$

The electrostatic origin of the centripetal force, along with Eqs. (14) and (16) then yields the Bohr states energies

$$E_n \approx mc^2 - me^4 \left(2n_\varphi^2 \hbar^2\right)^{-1}, \tag{17}$$

and differences between the state energies then gives the hydrogen spectra. The procedure of including the particle's self-field, along with quantized flux, thus leads to the standard (observed) hydrogen spectra of atomic physics.

Equation (14), upon identifying $2\pi R n_\varphi^{-1}$ as the "wavelength", can be seen to be a version of the de Broglie relationship. The version here is intimately connected to the presence of path curvature (while the conventional de Broglie relationship is claimed to be valid for a free particle). Consequently there is a difference between conventional quantum mechanics and the self-field/quantized flux approach taken here.

*All* particle "interference" experiments have at least one particle path being a non-straight line, thus the curvature form of Eq. (14) could represent the greater validity. If so, then "wave-particle" duality might be ultimately understood in terms of a particle and its self-field.

Here lack of a spot on a screen might be the result of the lack of a particle being present, rather than destructive interference between particle-waves. An electron propagating from cathode to diffraction grating, and then to an anode screen, must have an electrical return path. Thus there is an (are many) allowed closed path(s) which satisfy Eq. (13).

## ORBITS IN THREE DIMENSIONS

The concept of self-field is extended, in an obvious way, to the angular momentum

$$\boldsymbol{j}^{(s)} = qc^{-1}\,[r\mathrm{x}(A_\varphi^{(s)}\vec{\varphi}) + r\mathrm{x}(A_\theta^{(s)}\vec{\theta})]\text{, or} \tag{18}$$

$$\boldsymbol{j}^{(s)} = qc^{-1}[-rA_\varphi^{(s)}\vec{\theta} + rA_\theta^{(s)}\vec{\varphi}] \tag{19}$$

where $\vec{\varphi}$ and $\vec{\theta}$ are unit vectors. Again for circular orbits

$$2\pi R A_\theta^{(s)} = \Phi_\theta^{(s)}, \quad \Phi_\theta^{(s)} = n_\theta h c e^{-1}, \quad n_\theta = 0,1,2,3,... \tag{20}$$

Eqs. (11), (13), (19), and (20) then combine to give

$$\boldsymbol{j}^{(s)} = q(2\pi c)^{-1}[n_\varphi^{(s)}\vec{\theta} + n_\theta^{(s)}\vec{\varphi}]\,hce^{-1}. \tag{21}$$

Consider

$$J = j^{(p)} + j^{(s)} , \qquad (22)$$

with *J* being the total angular momentum. The existence of s-states then implies that

$$J = [-\left(n_\varphi^{(s)} - n_\varphi^{(p)}\right)\vec{\theta} + \left(n_\theta^{(s)} - n_\theta^{(p)}\right)\vec{\varphi}\,]\hbar , \qquad (23)$$

Where separate *n* indices have now been introduced for the particle and the self-field. For a single electron system we can choose the coordinate system at will. It is convenient to choose $n_\theta^{(s)} = n_\theta^{(p)} = 0$. With that choice

$$J = \left(n_\varphi^{(p)} - n_\varphi^{(s)}\right)\hbar\,\vec{\varphi} . \qquad (24)$$

Imagine an electron initially at infinity with $n_\varphi^{(p)} = n_\varphi^{(s)} = \infty$. It falls inward towards the nucleus by radiating a photon from the self-field, resulting in a p-state. Although the source of quantization is the flux, an imbalance between the self-field and particle can occur. Thus for a photon mediated transition

$$\Delta\left(n_\varphi^{(p)} - n_\varphi^{(s)}\right) = \pm 1 \qquad (25)$$

leading to the identification of the orbital quantum number $\ell$ as a measure of that imbalance.

$$\ell = n_\varphi^{(p)} - n_\varphi^{(s)} . \qquad (26)$$

Now turn on a magnetic field in the z-direction. $\vec{\theta}$ can be resolved into z and $\rho$ components. If $J \neq 0$ is always associated with a non-zero total magnetic moment, then the $\rho$ component of *J* is acted on by a magnetic torque, resulting in the generation of a $\vec{\varphi}$ component of *J* (the $\rho$ component precesses about the z-axis). The result, from Eq. (23), is

$$m = n_\theta^{(s)} - n_\theta^{(p)} \neq 0 . \qquad (27)$$

Because $n_\theta^{(s)}$ and $n_\theta^{(p)}$ are restricted to integers, m (not to be confused with mass) is also restricted to integers. Lastly, for $m = 0$ as the pre-external magnetic field condition, along with conserved *J*, we have

$$|m| \leq |\ell| . \qquad (28)$$

*m* is thus recognized as the magnetic quantum number of the more conventional quantum mechanics. Here $\ell = 0$ does not denote a zero angular momentum state *for the particle*, but rather zero angular momentum for the particle and self-field system.

**REFERENCES**

...
(1) M. A. Oliver, Found. Phys. Lett. **12**, 81 (1999); **11**, 61 (1998).
(2) A. O. Barut, *Electrodynamics and Classical Theory of Fields and Particles* (Dover, NY, 1980). Page 184.
(3) H. Goldstein, *Classical Mechanics* (Addison-Wesley, Reading, 1980). Page 322.
(4) L. I. Schiff, *Quantum Mechanics* (McGraw-Hill, NY, 1955). Second edition, page 320.